\documentclass[letterpaper, 10 pt, conference]{ieeeconf}  
\IEEEoverridecommandlockouts    
\usepackage{cite}
\usepackage{amsfonts}
\usepackage{algorithm}
\usepackage{algorithmicx}
\usepackage{textcomp}
\usepackage{float}
\usepackage{algpseudocode}
\usepackage{graphicx}
\usepackage{bm}
\usepackage{amsmath}
\usepackage{amssymb}
\usepackage{hyperref}
\usepackage{makecell}
\usepackage{xfrac}
\usepackage[margin=0.8in]{geometry}

\usepackage{color}
\usepackage{subcaption}

\newcommand{\Pfdroop}{\text{$P$-\,$\omega$ }}
\newcommand{\QVdroop}{\text{$Q$-$V$ }}
\newcommand{\KB}{\color{black}{}} 
\begin{document}



\title{\LARGE Optimizing Grid-Forming Controls using Relay-based Extremum Seeking to Enhance Transient Performance}

\author{Kyung-Bin Kwon$^{*}$, Min Gyung Yu$^{*}$, Sayak Mukherjee, Timothy I. Salsbury%
\thanks{Authors are with the Pacific Northwest National Laboratory, 902 Battelle Blvd, Richland WA 99354, $^*$ contributed equally.}%
\thanks{The research is supported by the E-COMP (Energy System Co-Design with Multiple Objectives and Power Electronics) Initiative at Pacific Northwest National Laboratory (PNNL), operated by Battelle for the U.S. Department of Energy under Contract DEAC05-76RL01830.}%
}
\maketitle

\begin{abstract}
Grid-forming (GFM) inverters are essential for enhancing stability in modern power systems with high penetration of inverter-based resources (IBRs). However, their performance highly depends on control parameters tuning, particularly the active power-frequency droop coefficient. This parameter presents a trade-off among competing objectives, including damping, settling time, rate of change of frequencies (RoCoF) and frequency nadirs. This paper proposes a real-time, adaptive optimization framework based on Extremum Seeking Control (ESC) to dynamically tune the GFM droop gain. A multi-objective cost function balances conflicting performance goals such as oscillation energy, frequency nadir, RoCoF, and post-disturbance settling performance. The approach is validated through numerical simulations on a modified IEEE 68-bus system. Results demonstrate that the cost function is convex with respect to the droop parameter, justifying gradient-based optimization. Furthermore, the ESC algorithm successfully tracks the time-varying optimal droop coefficient in real-time as network conditions change, thereby ensuring robust and near-optimal system performance without requiring an analytical grid model.

\end{abstract}

\vspace{-0.4cm}
\section{Introduction}
Grid-forming (GFM) technology enhances grid stability by autonomously regulating voltage and frequency at transmission interconnection points. Among various GFM strategies, droop control has emerged as one of the most widely adopted approaches \cite{wei}. This method leverages the active power–frequency ($P$–$f$) and reactive power–voltage ($Q$–$V$) droop characteristics to coordinate power sharing during both steady-state and transient operating conditions. The synchronous generator–like behavior of GFMs has motivated grid operators to consider large-scale deployment of GFM-interfaced resources in future power systems \cite{lasseter2019grid}. Over the years, several GFM implementations have been developed, including virtual synchronous machines (VSMs) \cite{compare_vsm_droop}, droop-controlled GFMs \cite{weiDu_droop_comparison}, and virtual oscillator control (VOC) \cite{brian_voc}, among others. These technologies offer distinct advantages over grid-following inverters (GFLs), particularly in frequency regulation, inertia emulation, black-start capability, and standalone microgrid operation. 

Studies have also shown that appropriate tuning of inverter control parameters can enhance overall system stability \cite{Chatterjee_def_statcom}. In droop-controlled GFMs, the active power–frequency droop coefficient plays a key role, as it directly influences system eigenvalues and their damping characteristics \cite{weiDu_droop_comparison, vsg_droop_comparison}. Recent research \cite{kundu2019identifying, nandanoori2020distributed, gorbunov2021identification, siahaan2024decentralized} has focused on deriving analytical stability conditions by defining upper bounds on droop gains. However, most of these results are limited to microgrids and do not fully address the challenge of achieving sufficient damping in large-scale, heterogeneous bulk power systems. For large-scale systems, conducting theoretical stability analyses using detailed inverter models is highly complex and often analytically intractable. Consequently, practical methods for obtaining stability-informed design parameters typically depend on numerical simulations \cite{weiDu_droop_comparison, droop_small_signal}. \cite{hossainstability} presents different stability assessments, and a zeroth-order gradient descent-based droop design to optimize for small-signal stability margins, however that requires detailed model information to perform the parametric stability analysis. 

\textit{Contributions.} The contribution of this paper is three-fold:
\begin{itemize}
  \item 
  We develop an extremum seeking control (ESC) algorithm that autonomously adjusts the active power--frequency droop gain of GFM inverters in real-time without requiring an explicit system model. This enables online adaptation to varying conditions and improved robustness of GFM operation under uncertainty.
  
  \item 
  A unified cost function is proposed that integrates oscillation energy, rate-of-change-of-frequency (RoCoF), and settling characteristics. This formulation captures the inherent trade-offs between frequency nadir, damping, and recovery speed, thereby allowing principled optimization of the droop coefficient through ESC.

  \item 
  The proposed ESC-based droop controller is validated on a modified IEEE 68-bus system with multiple GFMs, demonstrating improved frequency nadir and damping compared with fixed-droop baselines. Additionally, guidelines for ESC parameter tuning (search bounds, dwell time, and step gain) are provided to facilitate real-world implementation.
\end{itemize}

\vspace{-0.2cm}
\section{GFM-Integrated Grid Model}

We consider a bulk power system consisting of $N$ buses, with $m$ synchronous generators (SGs) and $n$ grid-forming (GFM) inverters. The dynamics of each SG are represented by the classical swing equations \cite{kundur}:
\begin{align}
&\dot{\delta}_i = \omega_i - \omega_0,\\
&\dot{\omega}_i = \tfrac{1}{M_i}\left[D_i(\omega_0 - \omega_i) + P_i - P_{ei}\right],
\label{eq:sg}
\end{align}
where $\delta_i$ and $\omega_i$ denote the rotor angle and frequency of generator $i$, $P_i$ is the mechanical input power, and $P_{ei}$ is the electrical power output. The parameters $M_i$ and $D_i$ correspond to the inertia and damping coefficients, respectively. The grid-forming inverters (GFMs) are modeled as \Pfdroop– and \QVdroop–based droop-controlled resources, following the WECC-approved positive-sequence model $REGFM\_A1$ \cite{REGFM_A1}, which was originally developed at PNNL. Their dynamic behavior can be described as \cite{REGFM_A1}:
\begin{subequations}
\label{eq:gfm}\vspace{-0.2cm}
\begin{align}
&\dot{\delta}j = \omega_j - \omega_0,\\
&\dot{\omega}_j = \frac{1}{\tau_j}\left[\omega_0 - \omega_j + m{p_j}(P_j^{set} - P_j)\right], \label{omega_eq}\\
&\dot{V}^e_j = \frac{1}{\tau_j}\left[V_j^{set} - V_j - V^e_j + m{q_j}(Q_j^{set} - Q_j)\right],\\
&\dot{E}_j = k^{pv}_j \dot{V}^e_j + k^{iv}_j V^e_j ,
\end{align}
\end{subequations}
%
where $\delta_j$ is the voltage angle, $\omega_j$ the frequency, $V_j$ the terminal voltage magnitude, and $E_j$ the internal voltage magnitude. The terms $P_j^{set}, Q_j^{set}, V_j^{set}$ denote the active-power, reactive-power, and voltage setpoints, while $P_j$ and $Q_j$ are the corresponding injections at the bus where GFM $j$ is connected. The auxiliary state $V^e_j$ captures the voltage error. Control parameters $m_{p_j}$ and $m_{q_j}$ represent the \Pfdroop and \QVdroop droop coefficients, respectively. $\tau_j$ is the measurement filter time constant, here chosen as $0.01$ s. Finally, $k^{pv}_j$ and $k^{iv}_j$ are the proportional and integral gains associated with the \QVdroop controller. Next, the active and reactive power balance at any bus $j,\; j=1,\dots,m$ can be expressed as,
\begin{subequations}
\label{eqn:load_flows}
\begin{align}
\label{eqn:load_active} 
0 &= P_{ej} - {\rm Re}\left\{ \sum\limits_{k = 1,\,k \ne j}^N V_j \left( V_{jk} B_{jk} \right)^* \right\} - V_j^2 G_j,\\
\label{eqn:load_reactive4} 
0 &= Q_{ej} - {\rm Im}\left\{ \sum\limits_{k = 1,\,k \ne j}^N V_j \left( V_{jk} B_{jk} \right)^* \right\} - V_j^2 B_j,
\end{align}
\end{subequations}
where $G_j$ and $B_j$ denote the conductance and susceptance of the shunt load at bus $j$ (including line charging). For lossless transmission lines, $B_{jk}$ denotes the susceptance of the tie-line between buses $j$ and $k$. In compact form, the power flow equations are written as $0 = g(x_s, x_f, V)$.

\section{Multi-Objective Dynamic Performance Criterion}

System dynamics vary significantly with different droop coefficients, affecting frequency nadirs, settling time, damping in nonuniform ways \cite{hossainstability}. 
This inherent trade-off necessitates an optimization approach to determine a droop coefficient that balances multiple objectives such as minimizing the frequency nadir, ensuring fast settling, and maintaining adequate damping via nonlinear oscillation energy, under changing grid operating conditions. To evaluate and minimize the overall system performance under droop control, a total cost function $J_{total}$ is defined. The objective is to find the optimal droop coefficient that balances frequency stability and energy efficiency. The cost function integrates multiple performance metrics that capture different aspects of system dynamics during frequency disturbances.

\subsubsection*{Average Energy Flow ($E_{\mathrm{avg}}$)}
This term captures the net energy exchanged by the system during frequency deviation events, representing the energy impact of frequency fluctuations. It is calculated as the average energy consumed or produced per second over the analysis window, where a lower value implies that the system requires less corrective energy to maintain frequency stability. This is desirable for economic operation and indicates a well-damped response. The average energy flow is mathematically defined as:
\begin{equation}
    E^{\mathrm{avg}} = \frac{1}{T} \sum_{i=1}^{N} \frac{v_{\mathrm{inertia}}}{2} \Delta f(t_i)^2
\end{equation}
where $T$ is the total duration of the analysis window, $N$ is the number of discrete time samples, $v_{\mathrm{inertia}}$ is the virtual inertia parameter, and $\Delta f(t_i) = f(t_i) - f_{\mathrm{nom}}$ is the frequency deviation from the nominal frequency at $t_i$.

\subsubsection*{Average and Maximum Rate of Change of Frequency ($R_{\mathrm{mean}}$, $R_{\mathrm{max}}$)}
The Rate of Change of Frequency (RoCoF) quantifies how quickly the system frequency responds to a disturbance. We consider two metrics:
\begin{itemize}
    \item The \textbf{average RoCoF ($R^{\mathrm{mean}}$)} measures the typical magnitude of frequency change rate over the time window.
    \item The \textbf{maximum RoCoF ($R^{\mathrm{max}}$)} captures the single most severe transient event during the window.
\end{itemize}
Smaller RoCoF values indicate smoother and more stable dynamic behavior, thereby reducing mechanical stress on synchronous generators and minimizing the risk of unnecessary protection relay activations. These metrics are calculated from discrete frequency measurements as follows:
\begin{equation}
    R^{\mathrm{mean}} = \frac{1}{N-1} \sum_{i=1}^{N-1} \left| \frac{f(t_{i+1}) - f(t_i)}{\Delta t} \right|
\end{equation}
\begin{equation}
    R^{\mathrm{max}} = \max_{i} \left( \left| \frac{f(t_{i+1}) - f(t_i)}{\Delta t} \right| \right)
\end{equation}
where $f(t_i)$ is the frequency at time step $i$, and $\Delta t$ is the time interval between samples.

\subsubsection*{Final Fluctuation Height ($F^{\mathrm{final}}$)}
This metric evaluates the system's ability to return to a steady state after a disturbance. Instead of measuring the total time to settle, it quantifies the magnitude of the largest frequency deviation remaining near the end of the analysis window (e.g., the final 20\% of the time period). A smaller final fluctuation height indicates that the system oscillations are well-damped and the frequency is stabilizing close to its nominal value. This serves as a practical proxy for assessing settling performance. It is defined as the maximum absolute frequency deviation in the final portion of the time window:
\begin{equation}
    F^{\mathrm{final}} = \max_{t \in [T_{\mathrm{final\_start}}, T_{\mathrm{end}}]} \left| f(t) - f_{\mathrm{nom}} \right|
\end{equation}
where $[T_{\mathrm{final\_start}}, T_{\mathrm{end}}]$ represents the final segment of the analysis window. A lower $F^{\mathrm{final}}$ is a key indicator of effective frequency control and system stability.

\subsubsection*{Objective Function}
The total cost is computed as a weighted sum of these performance indices. Each weight represents the relative importance assigned to the corresponding metric, depending on system design priorities and operational considerations. The optimization problem is formulated for the considered time period $[t_1, t_2]$ as:
\begin{align}
\label{eq:joint_cost}
    \min_{\mathbf{x}(\tau), \tau \in [t_1, t_2]} &J_{\mathrm{total}}(t)
    = w_1 E^{\mathrm{avg}}_{[t_1, t_2]}(\mathbf{x}) + 
    w_2 R^{\mathrm{mean}}_{[t_1, t_2]}(\mathbf{x}) \nonumber
    \\
    &+ w_3 R^{\mathrm{max}}_{[t_1, t_2]}(\mathbf{x}) + 
    w_4 F^{\mathrm{final}}_{[t_1, t_2]}(\mathbf{x}),
\end{align}
where $\mathbf{x}$ denotes the controlled droop gain of the GFM inverters. 
{\KB The weight vector $[w_1, w_2, w_3, w_4]$ is a tunable design parameter encoding operator trade-offs between energy efficiency, transient performance, and final settling stability. To ensure comparability, each metric is normalized by a scenario-specific reference value drawn from the baseline simulations so that weights operate on a common, unitless scale. The heuristic weight set used in this study, [6, 1, 0.0015, 35], emphasizes steady-state frequency restoration while balancing average and peak transient measures. These weights are intentional design choices and may be adjusted to reflect local priorities and operational limits. The corresponding sensitivity analysis yields absolute sensitivities of [0.31, 0.76, 0.27, 0.76], indicating robust tuning trends under reasonable weight variations.}

$J_{\mathrm{total}}(t)$ arises from nonlinear closed-loop dynamics and is therefore not globally convex. Since ESC generally requires convexity for guaranteed convergence, its local suitability for ESC will be evaluated numerically in Section~\ref{sec:result}. 

\section{ESC-based Adaptive GFM Gain Optimization}

\begin{algorithm}[b!]
\caption{Pseudo code implementation of the ESC-based real-time optimization algorithm}
\label{code:alg}
\begin{algorithmic}[1]
\State \textbf{Inputs:} Grid transient frequency trajectories, $\epsilon_t = 1$, $g_t = 0$, $d_t = 0$, $d_{lim} = 1$, Cost computation framework for $J$ as in \eqref{eq:joint_cost}.
\State $K = \Delta t(\mathbf{x}_{max}-\mathbf{x}_{min})/ 10$
\While{$t > 0$}
  \State $g_t = J'_t - J'_{t-1}$ 
  \If{$g_t > 0$ \textbf{and} $d_t \geq d_{lim}$}
    \State $\epsilon_t = -\epsilon_{t-1}, d_t = 0$ 
  \EndIf\\
  \;\;\; \textit{Droop Optimizer search:}
  \State $\mathbf{x}_t = \mathbf{x}_{t-1} + \epsilon_t * K$
  \State $\mathbf{x}_t = \max(\mathbf{x}_t, \mathbf{x}_{min}) \; \text{or} \min(\mathbf{x}_t, \mathbf{x}_{max})$
  \State $d_t = d_{t-1} + \Delta t$
\EndWhile
\end{algorithmic}
\end{algorithm}

\begin{figure}[t!]
    \centering
\includegraphics[width=0.4\textwidth]{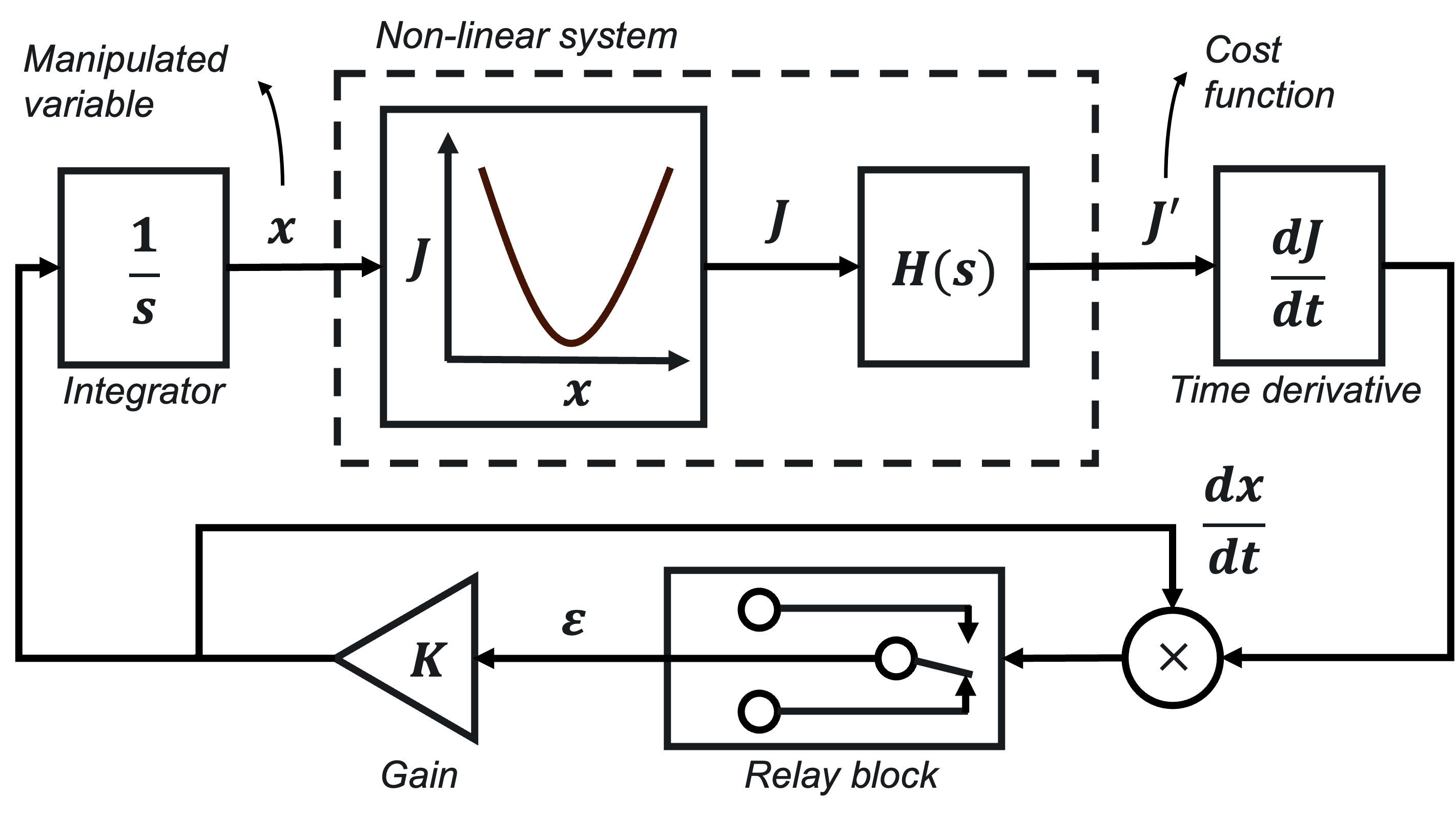}
    \vspace{-0.15 cm}
    \caption{Block diagram of the real-time ESC optimizer.}
    \label{fig:esc_block}
\vspace{-0.35cm}
\end{figure}

ESC approach operates as a feedback-driven mechanism grounded in the ``perturb and observe'' concept \cite{krstic2000stability}. It introduces a small perturbation to the system, monitors how the output responds, and then adjusts the control input accordingly to move toward optimal performance. In this study, we employ a single-input, single-output (SISO) relay-based ESC scheme, illustrated in Figure~\ref{fig:esc_block} (see \cite{salsbury2023}). {\KB Note that each iteration of the ESC corresponds to a complete disturbance event (e.g., a load change or line trip) over a time window sufficient to compute the metrics in the cost function. In real-time implementation, the ESC takes a step toward a more stable droop setting following each individual disturbance, though multiple disturbances are typically required for full convergence to the optimal value. Alternatively, the gain tuning strategy can be deployed on a digital twin of the real system, enabling optimal droop gains to be determined offline and then implemented directly in the physical system in anticipation of network topology or operating condition changes.}

In Figure~\ref{fig:esc_block}, the relay block produces a binary output $\epsilon \in {+1,-1}$, determined by the sign of the gradient of the cost function $J'$, and is designed to move along the direction of the negative gradient. To prevent excessive switching and allow adequate system response, time-scale separation is applied through a dwell time ($d_{lim}$, set to one sample in these experiments). The resulting relay output $\epsilon$ is scaled by a gain factor $K$ and integrated to generate the manipulated variable $x$, which is constrained within predefined bounds. The gain factor $K$ is defined as:
\begin{equation}
K = \frac{\Delta t (x_{max} - x_{min})}{10}
\end{equation}
\noindent where $\Delta t$ is the discrete time step and the value in the denominator governs the step size in $x$ per iteration - in this case 10\% of the specified search range $\{x_{max},x_{min}\}$. In this work, the manipulated variable $x$ represents the per droop parameter of the GFM inverters.

\begin{figure}[t!]
    \centering
\includegraphics[width=0.35\textwidth]{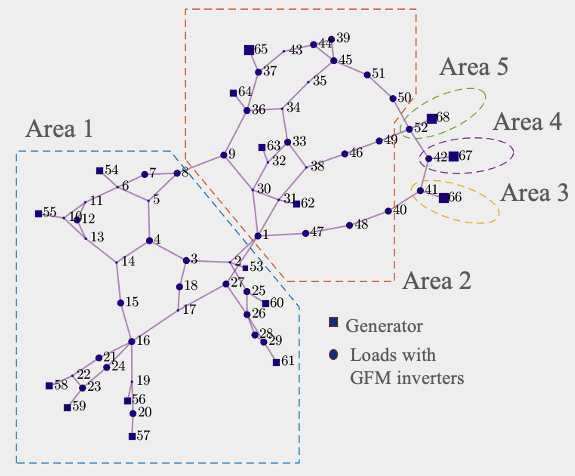}
    \caption{IEEE 68-bus system with 35 GFM inverters.}
    \label{fig:68bus}
    \vspace{-0.35cm}
\end{figure}

\section{Numerical Testing}
\label{sec:result}
For these tests, we consider the positive-sequence model of the IEEE 16-machine, 68-bus test system as depicted in Figure~\ref{fig:68bus} \cite{hossainstability}. The generator and network parameters are obtained from the standard data set, with modifications to lower the inertia. The test system is modified to include GFM IBRs on select buses. There are a total of 35 such inverters, each at a load bus --- note that the inverter buses are represented by the set $\mathcal{N}_{I}$.


To evaluate the proposed cost function and its relationship with the droop coefficient, four different simulation scenarios were considered:

\begin{itemize}
    \item \textbf{Scenario 1:} Baseline network configuration (all branches operational).
    \item \textbf{Scenario 2:} One transmission branch \textit{(Bus 26-28)} removed (updated $Y_{\mathrm{bus}}$ matrix).
    \item \textbf{Scenario 3:} Two transmission branches \textit{(Bus 26-28, 41-42)} removed (updated $Y_{\mathrm{bus}}$ matrix).
    \item \textbf{Scenario 4:} One transmission branch \textit{(Bus 25-26)} removed (updated $Y_{\mathrm{bus}}$ matrix).
\end{itemize}

\begin{figure}[!t]
    \centering
    \begin{subfigure}{0.23\textwidth}
        \centering
        \includegraphics[width=\textwidth]{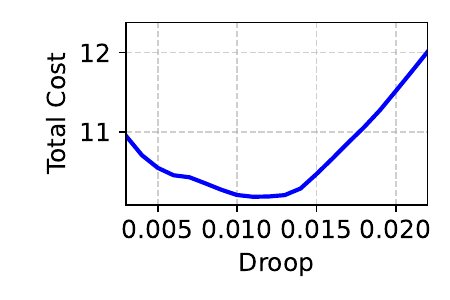}
            \vspace{-0.5 cm}
        \caption{Scenario 1 (no branch removed)}
        \label{fig:convexity_test_Ybus0}
    \end{subfigure}
    \hfill
    \begin{subfigure}{0.23\textwidth}
        \centering
        \includegraphics[width=\textwidth]{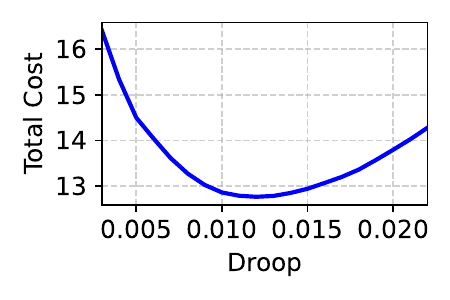}
            \vspace{-0.5 cm}
        \caption{Scenario 2 (Bus 26-28 removed)}
        \label{fig:convexity_test_Ybus1}
    \end{subfigure}
    
    
    \begin{subfigure}{0.23\textwidth}
        \centering
        \includegraphics[width=\textwidth]{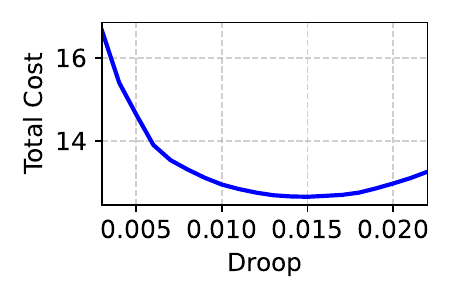}
            \vspace{-0.5 cm}
        \caption{Scenario 3 (bus 26-28, 41-42 removed)}
        \label{fig:convexity_test_Ybus2}
    \end{subfigure}
    \hfill
    \begin{subfigure}{0.23\textwidth}
        \centering
        \includegraphics[width=\textwidth]{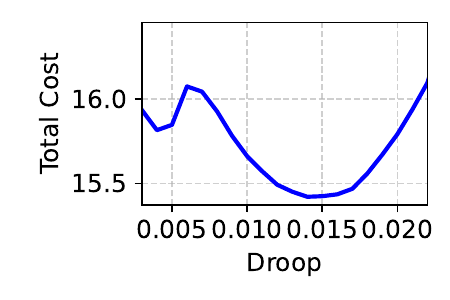}
            \vspace{-0.5 cm}
        \caption{Scenario 4 (bus 25-26 removed)}
        \label{fig:convexity_test_Ybus1_seed1}
    \end{subfigure}

    \caption{Convexity tests across network configurations.}
    \label{fig:convexity_tests_all}
    \vspace{-0.65cm}
\end{figure}

Figures~\ref{fig:convexity_tests_all} illustrate the total weighted cost as a function of the per-unit droop coefficient for all scenarios. The results describe the sensitivity of the total cost to the changes in the droop coefficient, and this sensitivity varies with system topology. In general, the observed shape of the cost curve with respect to the droop parameter (ranging from 0.003 to 0.022) is a convex shape.
\color{black}
While this convexity is shown here for analysis, such cost curve is generally unknown to the system designer in a real-world application. This lack of a-priori knowledge makes ESC a well-suited, model-free approach for identifying the optimal operating points online.
\color{black}
\color{black}

In Scenario~1, the minimum total cost occurs at a droop of approximately 0.011, with cost values ranging from 10 to 12. In contrast, Scenarios~2–4 exhibit higher overall cost levels (approximately 13–17). The optimal droop shifts to around 0.012 in Scenario~2, and 0.015 in Scenario~3. Scenario~4 shows a less distinct convex pattern. It shows a local convex near 0.003–0.005, but the overall minimum cost occurs at a droop of approximately 0.014. This behavior indicates that both the location and shape of the optimal region are influenced by the system topology, as reflected in changes to the $Y_{\mathrm{bus}}$ matrix.
\begin{figure}[!t]
    \centering
    \begin{subfigure}{0.45\textwidth}
        \centering
        \includegraphics[width=\textwidth]{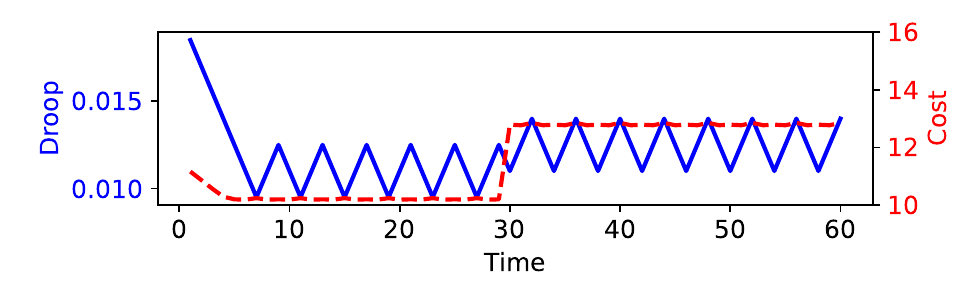} 
            \vspace{-0.7 cm}
        \caption{case1}
        \label{fig:ESC_droop_Ybus_01_seed139}
    \end{subfigure}
    \begin{subfigure}{0.45\textwidth}
        \centering
        \includegraphics[width=\textwidth]{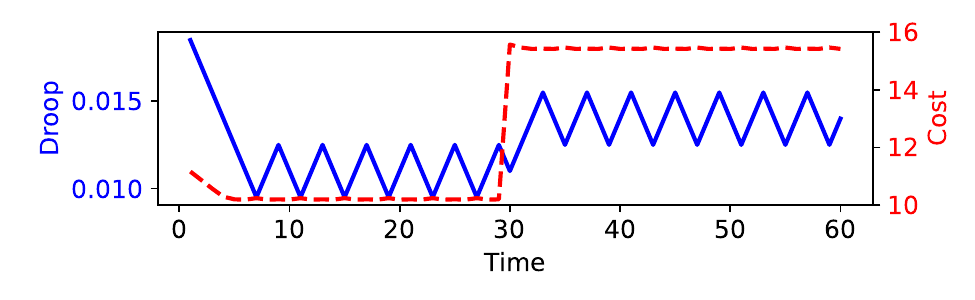}
            \vspace{-0.7 cm}
        \caption{case2} 
        \label{fig:ESC_droop_Ybus_01_seed1}
    \end{subfigure}
    \begin{subfigure}{0.45\textwidth}
        \centering
        \includegraphics[width=\textwidth]{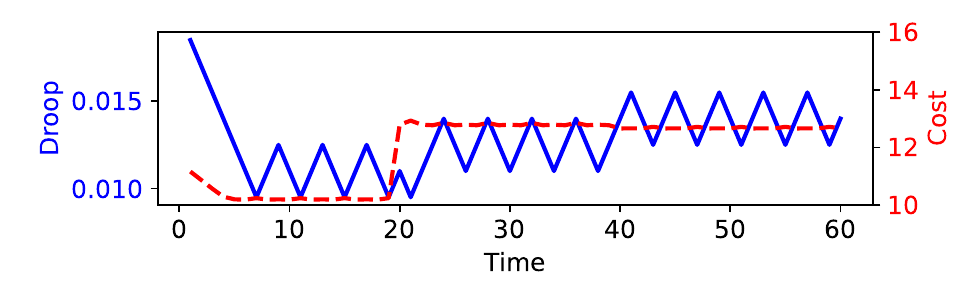}
            \vspace{-0.7 cm}
        \caption{case 3} 
        \label{fig:ESC_droop_Ybus_012}
    \end{subfigure}
    \caption{Dynamic performance of ESC: the droop value (left axis) and corresponding total cost (right axis) under switching scenarios showing real time adaptation.}
    \label{fig:ESC_droop_cost}
    \vspace{-0.65cm}
\end{figure}
To further evaluate ESC’s ability to adapt to changing conditions, additional simulations were performed where transmission branches were dynamically removed from the baseline network (Scenario~1) during operation. Three test cases were considered:
\begin{itemize}
\item \textbf{Case 1:} Scenario1($t\!\in\![0,30]$)$\rightarrow$Scenario2($t\!\in\![30,60]$).
\item \textbf{Case 2:} Scenario1($t\!\in\![0,30]$)$\rightarrow$Scenario4($t\!\in\![30,60]$).
\item \textbf{Case 3:} Scenario1($t\!\in\![0,20]$) $\rightarrow$ Scenario2($t\!\in\![20,40]$) $\rightarrow$ Scenario3($t\!\in\![40,60]$).
\end{itemize}

\begin{figure}[t!]
    \centering
    \begin{subfigure}{0.495\columnwidth} 
        \centering
        \includegraphics[width=\linewidth]{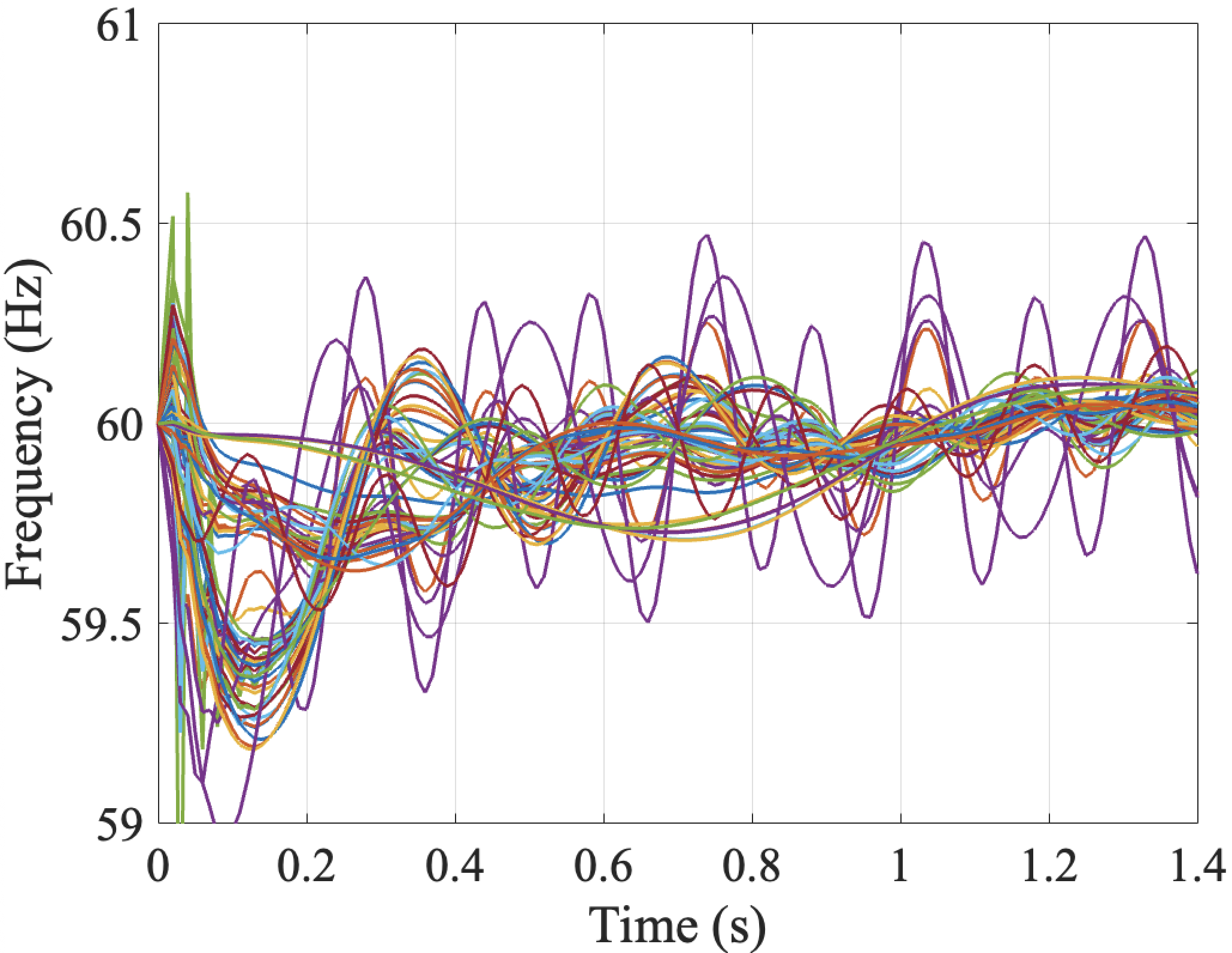}
        \caption{} 
        \label{fig:test_esc}
    \end{subfigure}\hfill 
    \begin{subfigure}{0.495\columnwidth} 
        \centering
        \includegraphics[width=\linewidth]{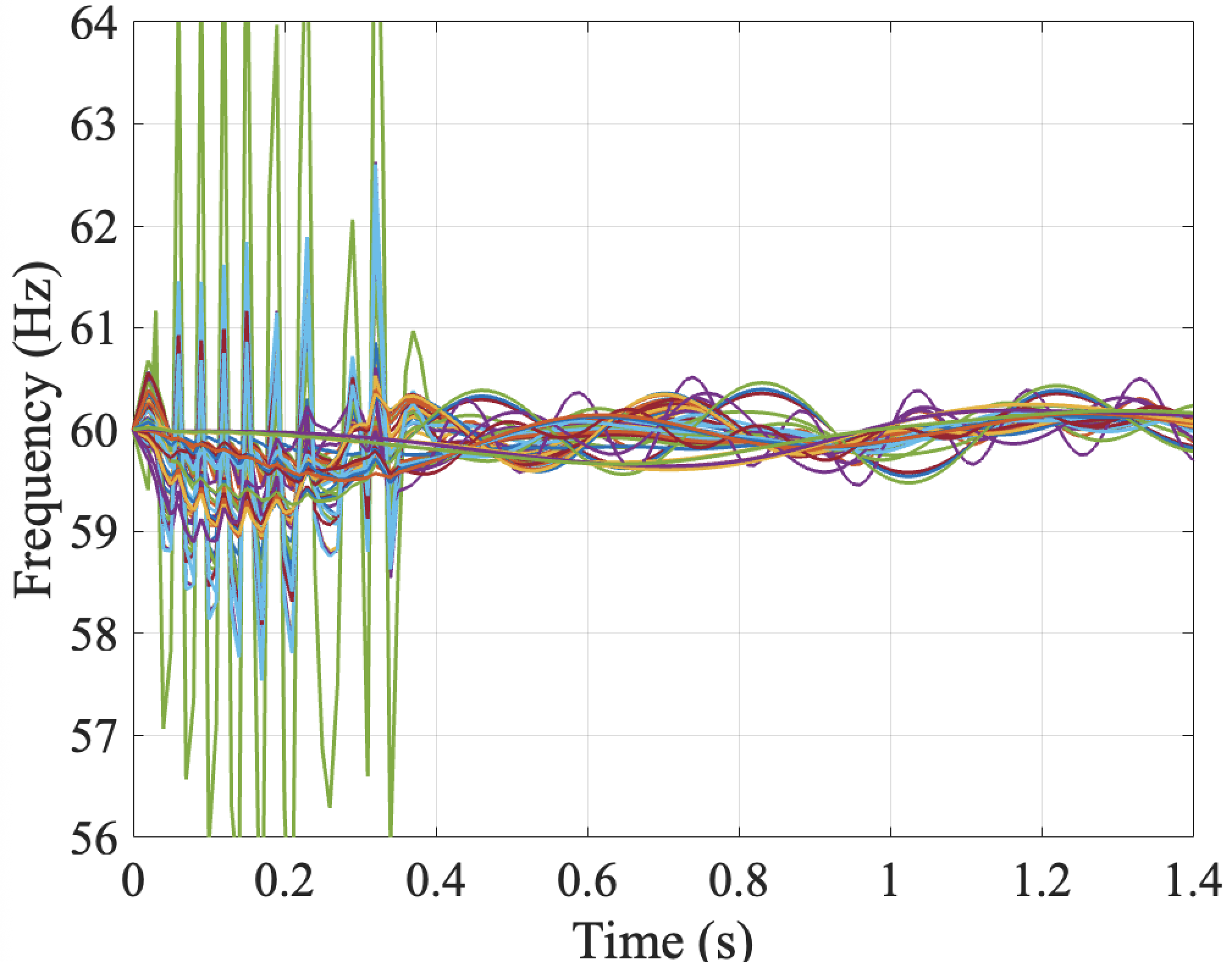}
        \caption{} 
        \label{fig:test_base}
    \end{subfigure}
    \caption{Frequency deviation of Scenario~1 when droop coefficient is set at (a) ESC-controlled value ($0.0125$) and (b) a heuristic base value ($0.0235$).}
    \label{fig:test}
    \vspace{-0.55cm}
\end{figure}

Figure~\ref{fig:ESC_droop_cost} illustrates the dynamic response of the ESC controller under the three switching cases. In Case~1, during the first interval (Scenario~1), the ESC oscillates around 0.010–0.012, matching the static optimal droop of 0.011. After the system change (Scenario~2), it adapts to 0.012–0.014, again aligning with the new optimum.
In Case~2, when the system transitions to Scenario~4, the ESC shifts to approximately 0.0125–0.015, consistent with the previously identified optimal point.
In Case~3, the system topology changes more dynamically, first to Scenario~2 and later to Scenario~3. Although the overall cost variation is modest between Scenario~2 and Scenario~3, the optimal droop shifts slightly toward 0.015, and the ESC successfully follows this trend. 
These results confirm that the ESC algorithm effectively tracks the time-varying optimal droop coefficient and maintains near-optimal performance despite changing network conditions. Note that one iteration of the ESC corresponds to a completed load change over a time window long enough to compute the metrics used in the cost function. 
\color{black} It is crucial to recognize that this persistent oscillation is a core feature of the ESC. This small, persistent fluctuation in the droop gain is the necessary trade-off for the algorithm's inherent adaptability. Recent research focuses on mitigating this trade-off by using adaptive gains to reduce the oscillation amplitude as the system converges to the optimum~\cite{salsbury2020self, salsbury2024new}. 
Lastly, Figure~\ref{fig:test} compares the system's frequency response with the ESC-controlled droop coefficient ($0.0125$) against a heuristically tuned base value ($0.0235$). 
As shown in the figure, the ESC method leads to a stable response where the initial frequency oscillations following a disturbance at $t=0$ are effectively damped, though both the line trip and huge load changes happen at the same time.
On the other hand, the base case results in severe and sustained oscillations, with frequency deviations far outside acceptable operational limits. 
This clearly demonstrates that the proposed ESC method provides significantly better transient stability.

\color{black}

\section{Paper Summary \& Future Directions}
This paper addressed the critical challenge of optimally tuning the active power-frequency droop coefficient for grid-forming (GFM) inverters in bulk power systems.  To resolve the tradeoff in droop selection for multiple dynamic performance criterion, we proposed a model-free, adaptive optimization framework based on Extremum Seeking Control (ESC). A multi-objective cost function was designed to holistically capture system dynamic performance, and numerical tests on a modified IEEE 68-bus system confirmed its close-to-convex shape with respect to the droop gain, validating the suitability of ESC, and subsequent dynamic simulations demonstrated that the algorithm effectively and autonomously tracks the optimal droop gain in real-time, adapting successfully to significant changes in system topology. Thus, this work presents a practical and robust method for enhancing the dynamic performance of GFM-integrated grids without reliance on complex analytical models. {\KB Future work includes extending the framework to a multi-input, multi-output (MIMO) ESC to co-optimize active and reactive power droop gains, developing systematic methods for cost-function weight selection, incorporating more detailed synchronous generator dynamics, and enabling spatially-varying droop coefficients to exploit disturbance locality.}

\bibliographystyle{IEEEtran}
\bibliography{refs}

\end{document}